\documentstyle[12pt]{article}
\begin{document}
\begin{center}
\LARGE
Polymerization in a Ferromagnetic Spin Model with Threshold
~\\
~\\
\vskip 3. truecm
\normalsize
Emilio N. M. Cirillo and Sebastiano Stramaglia\\
{\it Dipartimento di Fisica dell'Universit\`{a} di Bari} and\\
{\it Istituto Nazionale di Fisica Nucleare, Sezione di Bari\\
via Amendola 173, 70126 Bari, Italy}\\
E\_mail: cirillo@axpba0.ba.infn.it, stramaglia@axpba5.ba.infn.it\\
~\\
~\\
\end{center}
\vskip 3.5 cm
\begin{abstract}
We propose a spin model with a new kind of ferromagnetic interaction, which may 
be called {\it ferromagnetic coupling with threshold}. In this model the 
contribute of a given spin to the total energy has only two possible values and 
depends on the number of parallel spins among its nearest and next to the 
nearest neighbors. By mapping the model onto the Ising version of the 
isotropic eight vertex model, we obtain some evidence of a low temperature 
phase made of alternate parallel pluses and minuses polymers. 
\end{abstract}

\newpage
\par\noindent
{\bf 1. Introduction.}
\addtolength{\baselineskip}{\baselineskip}
\vskip 0.5 truecm
\par\noindent
Ising--like spin systems are deeply studied statistical models which have a 
surprising richness of critical behaviour (see, e. g., \cite{BAX}).
The axial next nearest neighbor Ising (ANNNI) model is the simplest model which 
describes modulated structures; experimental evidences for such phases are 
furnished by binary alloys (see, e. g., \cite{YEO} and \cite{SEL}). Modulated 
ordering in ANNNI model is the effect of the competition between ferromagnetic 
couplings (between nearest neighbor spins) and antiferromagnetic couplings
(between second--neighbor spins along one lattice direction). This competition 
leads to series of commensurate and incommensurate modulated phases of 
arbitrarily long wavelength (see, e. g., \cite{FS}).
\par
In this paper we show that a system where the spins tend to be 
aligned to the majority of neighbor spins has a low temperature striped phase;
this phenomenon is not pointed out by the ANNNI model and might
lend some insight to the problem of the formation of striped--like patterns.
\par
The model we propose here is an Ising model where the contribution of a given 
spin to the hamiltonian is $-\beta$ if the spin is aligned to the majority of 
its neighbors and $+\beta$ if it is not.
In the following we give the definition of the model that we are going to 
discuss. Let us consider the infinite bidimensional square lattice 
$\Lambda = {\bf Z}^{\bf 2}$ and define a ${\bf Z}_{\bf 2}$ spin variable 
$s_i\in\{-1,+1\}$ on each site $i\in\Lambda$. We denote by 
$s\in\Omega =\{-1,+1\}^{\Lambda}$ a configuration of the system and define the
formal hamiltonian
\begin{equation}
H(s)=-\beta\; \sum_{i\in\Lambda}s_i\phi_i(s)\;\; ,
\label{eq:ham}
\end{equation}
where $\beta$ is a positive real number and
\begin{equation}
\phi_i(s)={\rm sign}\ \{\sum_{j=1}^9 s_{i,j}\}\;\;\forall i\in\Lambda ,
\label{eq:maj}
\end{equation}
where we have denoted by $s_{i,j}\; j=1,...,9$ the nine spins in the 
$3\times 3$ squared block $B_i$ centered on the site $i$. All the equilibrium 
properties of the model can be obtained from the partition function 
$Z_{\Lambda}(\beta)=\sum_{s\in\Omega} {\rm exp} (-H(s))$.
\par  
We remark that $\beta >0$ induces a sort of ferromagnetic coupling among the 
spin variables, but model (\ref{eq:ham}) is ferromagnetic in a different 
fashion with respect to the Ising model $H_{I}=-\beta\sum_{<i,j>}s_is_j$, where 
the sum runs over the pairs of nearest neighbors sites in $\Lambda$. We call 
this kind of coupling {\it ferromagnetic coupling with threshold}.
\par
Indeed, let us suppose $s_i=+1$: in the Ising model the contribute of the spin 
$s_i$ to the total energy of the system decreases when the number of nearest 
neighbor plus spins increases, hence by increasing the number of plus spins 
among $s_i$ nearest neighbors we get more and more energetically preferable 
situations. In our model the contribute of the spin $s_i$ to the total energy 
has only two possible values: $+\beta$ if the number of plus spins in $B_i$ is 
less than five, $-\beta$ if it is greater than or equal to five; hence, from 
the point of view of the spin $s_i$, the best situation is reached when four of 
the eight spins in $B_i\setminus\{i\}$ are equal to $+1$ and there is no 
further energetic gain if the number of plus spins in $B_i\setminus\{i\}$ 
increases up to eight.
\par
As it will be explained below, due to the freedom in the choice of number and 
location of the plus (minus) spins around a fixed plus (minus) one, our model 
shows a very peculiar behaviour at low temperatures. In the following 
we give some evidence of a critical phase transition to a low temperature 
phase, which is a sort of ``polymer phase" made of alternate stripes with 
fluctuating boundaries.
\par
The occurrence of a striped phase does not depend on the geometry of 
the blocks $B_i$, which appear in the definition of model (\ref{eq:ham}). 
We find a striped phase also for the model  
\begin{equation}
{\tilde H}(s)=-\beta\; \sum_{i\in\Lambda}s_i{\tilde \phi}_i(s)\;\;\forall
s\in\Omega \;\; ,
\label{eq:hamti}
\end{equation}
where $\beta$ is the inverse temperature and
\begin{equation}
{\tilde \phi}_i(s)={\rm sign}\ \{\sum_{j=1}^5 s_{i,j}\}\;\;\forall 
i\in\Lambda\;\; ,
\label{eq:majti}
\end{equation}
where we have denoted by $s_{i,j}\; j=1,...,5$ the five spins in the ``cross" 
block $C_i$ containing the site $i$ and its four nearest neighbors.
\par
The hamiltonian (\ref{eq:ham}) is not strictly ferromagnetic, we mean that it 
does not satisfy the conditions under which Griffiths's inequalities hold 
\cite{GRIF}. 
Indeed by expanding $H$ as a linear combination of products of the spin 
variables it can be shown that in model (\ref{eq:ham}) the couplings between 
nearest and next to the nearest neighbor spins are ferromagnetic, but 
antiferromagnetic terms are present, as well; in the case of model 
(\ref{eq:hamti}) the coupling between nearest neighbors is ferromagnetic, while 
the coupling between next to the nearest neighbors is antiferromagnetic.
In both the two models a competition between ferromagnetic and 
antiferromagnetic couplings occurs, this is similar to what happen in the case 
of the ANNNI model.
\par
Due to this competition it is not surprising to find striped 
phases in our model. We remark that this competition is essentially 
 an ``entropic" effect. The spin $s_i$ prefers to have 
just four parallel spins in $B_i\setminus\{i\}$: this is energetically 
equivalent to the situation with eight parallel spins in $B_i\setminus\{i\}$, 
but the former can be realized in many different ways, while the latter just in 
one way.
 The phase transition has been studied by a 
Renormalization Group Transformation known as ``majority rule" (see, e. g.,
\cite{ma,NL1}); this is suggested by the structure of hamiltonian 
(\ref{eq:ham}).
\par
The model (\ref{eq:ham}) has also an interesting interpretation as a ``model of 
atoms", indeed, it can be seen as a model of a binary alloy, namely a mixture 
of two species of atoms (say A and B), such as the Ising model 
(see, e. g., \cite{KH}). But in the present case the configurations which are 
energetically preferred are those in which an A (B) atom has at least four A 
(B) atoms among its nearest and next to the nearest neighbors, no matter if the 
number of such atoms is greater than four. Due to this fact the ordered phase
is an alternate sequence of stripes of A and B atoms.
\par
The transformation (\ref{eq:maj}) is widely used for the 
recovering of noisy images and it is known as the {\it median filter} (see, e. 
g., \cite{TSH}). It follows that the hamiltonian (\ref{eq:ham}) provides a 
stochastic version of the median filter. 
\par
Finally, we observe that in our model each spin is coupled to a boolean 
function of its neighbors spins; this fact resembles the dynamical rules in 
the {\it random networks of automata}, where each spin is updated according to 
a random boolean function of its neighbors ({\it finite dimensional case})
or to a random boolean function of a certain number $k$ of random spins of the 
whole network ({\it mean field case}, see \cite{DER}). 
\par
Random network of automata exhibit dynamical phase transitions which can be 
studied by the ``distance method" (see, e. g., \cite{DER}). Our results
suggest that, in the finite dimensional case, equilibrium phase transitions may 
arise if one properly chooses the statistical ensamble of the boolean 
functions.
\par
The paper is organized as follows: in the next section the expansion of 
hamiltonians (\ref{eq:ham}) and (\ref{eq:hamti}) in terms of products of spin 
variables is performed. In section 3 the phase transition is studied by mapping 
the model onto the Ising version of the isotropic 8--vertex model. Section 4 
summarizes the conclusions.

%\newpage
\vskip 2 truecm
\par\noindent
{\bf 2. Some properties of the hamiltonian.}
\vskip 0.5 truecm
\par\noindent
In this section we discuss some properties of the model introduced above; in 
particular we describe its ground states, we give a rough evaluation of its
residual entropy and we expand its hamiltonian as sums of products of the 
spin variables.
\par
Model (\ref{eq:ham}) has an infinite number of ground states, which are the 
configurations satisfying the constraint 
\begin{equation}
s_i = \phi_i(s)\;\;\forall i\in\Lambda\;\; ; 
\label{eq:constr}
\end{equation}
such configurations are called 
{\it median roots}, that is configurations invariant under the 
median filtering. 
\par
A complete characterization of bidimensional median roots is still missing, 
we refer to \cite{TSH} for a related discussion and describe them 
heuristically: a median root is made of two typical kinds of structures, 
namely cluster--like structures and polymer--like structures.
In Fig. 1a we show the smallest cluster which is a root, a 12--spins cluster 
called {\it minroot} while in Fig. 1b a typical polymer structure is depicted.
\par
It follows that the model has a residual 
entropy; we obtain a rough estimate of it by considering the model defined on a 
cylinder of $6\times N$ sites with $N\to\infty$. On a $6\times N$ strip (with 
periodic conditions in the 6--sites direction) we can classify the 
median roots by the configurations on the last two columns. Taking into 
account 
the symmetries of the problem (parity $s\to -s$ and invariance under rotations 
of the cylinder) leads to $30$ not--equivalent classes. 
A $30\times 30$ transfer 
matrix $T_{a,b}$ can be defined as the number of ``a" roots on the $6\times (N+1
)$ cylinder obtained by any ``b" root on the $6\times N$ cylinder by adding a 
column. 
\par
The largest eigenvalue of the transfer matrix provides the residual 
entropy. We do not report here the details, but only quote the result. The 
largest eigenvalue is found to be $\lambda_0=3.57477$ and the residual entropy 
for site ${1\over 6}\ln\lambda_0 =0.21232$. To our knowledge there is not any 
estimate in literature for the entropy of the median roots to compare with 
ours.
\par
In order to expand the hamiltonian (\ref{eq:ham}) as
a linear combination of products of the spin variables we recall that if 
$f(s)$ is function of the ${\bf Z}_{\bf 2}$ spin variables $s_i$ 
where $i$ ranges over some finite set $V$, then $f(s)$ may be written in a 
unique way as
\begin{equation}
f(s)=\sum_{X}c(X)s(X)\;\; ,
\label{eq:four}
\end{equation}
where $s(X)=\prod_{i\in X}s_i$, the sum runs over all the 
subsets $X\subset V$ and the numbers $c(X)$ are given by
\begin{equation}
c(X)={1\over 2^{|V|}}\sum_{s\in\{-1,+1\}^{V}} s(X) f(s)\;\; ,
\label{eq:fourcoeff}
\end{equation}
where $|V|$ is the cardinality of $V$, that is the number of sites in $V$.
\par
Now, we observe that $H$ can be written as 
follows
\begin{equation}
H(s)=-\beta\sum_{i\in\Lambda} f_i(s)\;\; ,
\label{eq:hamff}
\end{equation}
where the function $f_i(s)=s_i\phi_i(s)$ is
defined on the finite sets $B_i$ $\forall i\in\Lambda$.
Hence, the function $f_i(s)$ can be expanded as in 
equation (\ref{eq:four}): one has to calculate $2^9-1=511$ coefficients (that 
is one has to consider all the possible subsets of $B_i$ excepted for the 
empty set). This number could be reduced taking properly into account the 
symmetries of the hamiltonian. 
\par
We have performed a computer assisted calculation of all the coefficients and, 
by working out the sums in equation (\ref{eq:hamff}), we have obtained
\begin{equation}
H(s)=-{\beta\over 2^9} \sum_{i=1}^{15} \gamma_i\sum_{\zeta\in\Gamma_i} 
s(\zeta)\;\; ,
\label{eq:hamex}
\end{equation}
where the families of clusters $\Gamma_i$ and the related coefficients 
$\gamma_i$ are given in Table 1. We remark that each $\Gamma_i$ represents a 
family of a certain kind of cluster of $\Lambda$ contained in a $3 \times 3$ 
block $B$; for example, $\Gamma_1$ is the family of all the pairs of nearest 
neighbor sites, $\Gamma_6$ is the family of all the plaquettes.
\par
This calculation shows that model (\ref{eq:ham}) is not ferromagnetic (see 
Table 1) in the sense that it fulfills the hypothesis under which 
Griffith's inequalities hold.
This model is characterized by ferromagnetic and antiferromagnetic couplings:
we observe that the 2--spins interactions are ferromagnetic if two adjacent 
columns or rows are involved (nearest and next to the nearest spins couplings),
while they are antiferromagnetic if the involved columns or rows are at 
distance two lattice spacings. As in the case of ANNNI model, one may expect 
that the result of this competition could be a low temperature striped--phase.
\par
By performing the same calculation in the case of model (\ref{eq:hamti}) we 
have obtained
$$
{\tilde H}(s)=-{\beta\over 2^5} \left[ 24\sum_{<i,j>}s_i s_j 
-8\sum_{<<i,j>>} s_i s_j\; +\right.
$$
\begin{equation}
\left.
-4\sum_{<<<i,j>>>}s_i s_j  -4\sum_{{\rm T}_{i,j,k,l}} s_i s_j s_k s_l
+ 12\sum_{\diamondsuit_{i,j,k,l}} s_i s_j s_k s_l\right]\;\; ,
\label{eq:hamexti}
\end{equation}
where the five sums run respectively over the pairs of nearest neighbors, the 
pairs of next to the nearest neighbors, the pairs of second neighbors along 
the lattice directions, the 4--sites cluster containing a site $i$ and 
three of its four nearest neighbors (T--shaped clusters), the square 4--sites 
clusters with sides at $45^o$ with respect to the lattice directions.
\par
Even model (\ref{eq:hamti}) exhibits a competition between ferromagnetic and 
antiferromagnetic couplings, hence we expect a low temperature striped phase in 
this case, as well.

%\newpage
\vskip 2 truecm
\par\noindent
{\bf 3. The phase transition.}
\vskip 0.5 truecm
\par\noindent
The structure of the hamiltonian (\ref{eq:ham}) suggests to introduce a 
new set of dynamical variables in order to investigate the phase diagram of 
our model.
\par
Suppose to partition $\Lambda$ into $3\times 3$ squared blocks $B_{\alpha}$ 
where $\alpha$ is the site that is in the centre of $B_{\alpha}$, the 
collection of all these sites $\alpha$ is denoted by $\Lambda'$. 
On each of these sites $\alpha$ we define the new variable 
\begin{equation}
s'_{\alpha}=\phi_{\alpha}\;\;\forall\alpha\in\Lambda' ;
\label{eq:rin}
\end{equation}
we have defined the new variables by ``integrating" over the fluctuations of 
the old ones on squared $3 \times 3$ blocks: we expect that the details of 
the configurations of the system on the scale of three lattice spacings are 
inessential to describe the peculiarities of the ordered phase.
\par
We have, then, introduced a new model defined on the lattice $\Lambda'$, with 
space of configurations $\Omega'=\{-1,+1\}^{\Lambda'}$; the equilibrium 
(unnormalized) measure of this new model is given by
\begin{equation}
\mu'(s')=\sum_{s\in\Omega} Z(s',s) e^{-H(s)}\;\; ,
\label{eq:rinorm}
\end{equation}
where it has been introduced the probability kernel 
$Z(s',s)=\prod_{\alpha\in\Lambda'} \delta_{s'_{\alpha},\phi_{\alpha}(s)}$. The 
formal hamiltonian of the new model is 
\begin{equation}
H'(s')=- {\rm log} \; \mu'(s')\ +\  {\rm const}\;\;\forall s'\in\Omega'.
\label{eq:newham}
\end{equation}
\par
In order to work out the sum in equation (\ref{eq:rinorm}) we use the method of 
the cumulant expansion (see \cite{NL}) writing the hamiltonian (\ref{eq:ham}) in
the form $H(s)=H_0(s)+V(s)\;\forall s\in\Omega$, where 
$H_0(s)=-\beta \sum_{\alpha\in\Lambda'} s_{\alpha}\phi_{\alpha}(s)$ contains 
the interactions between spins within the same block 
$B_{\alpha}\;\forall {\alpha}\in\Lambda'$, while 
$V(s)=-\beta \sum_{i\in\Lambda\setminus\Lambda'} s_i\phi_i(s)$ contains the 
interactions between spins belonging to different blocks. 
By truncating the cumulant expansion at the first order, one can show that the 
new hamiltonian is in the form
\begin{equation}
- H'(s')={\rm const}+J_{<>}\sum_{<\alpha\gamma>}s'_{\alpha}s'_{\gamma}
             +J_{<<>>}\sum_{<<\alpha\gamma>>}s'_{\alpha}s'_{\gamma}
	     +J_{
{\vcenter{\vbox{\hrule height.3pt
	    \hbox{\vrule width.3pt height0.15truecm \kern0.15truecm
	    \vrule width.3pt}
	    \hrule height.3pt}}}
}\sum_{_{\alpha} ^{\gamma}
{\vcenter{\vbox{\hrule height.3pt
	    \hbox{\vrule width.3pt height0.15truecm \kern0.15truecm
	    \vrule width.3pt}
	    \hrule height.3pt}}}
^{\delta}_{\eta}}
                s'_{\alpha}s'_{\gamma}s'_{\delta}s'_{\eta}\;\; ,
\label{eq:otto}
\end{equation}  
where the three sums run respectively over all the pairs of nearest neighbors, 
next to the nearest neighbors and over all the plaquettes in $\Lambda'$; we 
remark that the new model is the Ising version of the isotropic eight vertex 
model \cite{BAX}. With a computer assisted 
calculation we have obtained the new couplings $J_{<>},\; J_{<<>>}$ and 
$J_{
{\vcenter{\vbox{\hrule height.3pt
	    \hbox{\vrule width.3pt height0.15truecm \kern0.15truecm
	    \vrule width.3pt}
	    \hrule height.3pt}}}
}$ as functions of the original coupling $\beta$ (see Fig. 3); their 
structure is $J_{a}(\beta )=\beta F_{a}(e^{2\beta})$, where $F_{a}$ are 
rational functions.
\par
The variables transformation (\ref{eq:rin}) and the calculation of the 
hamiltonian of the new model amount to perform the ``majority rule" 
Renormalization Group Transformation (see \cite{ma,NL1}).
We remark that for $\beta$ large enough, with a single step of renormalization, 
the model (\ref{eq:ham}) is mapped into the eight vertex model 
(\ref{eq:otto}) with ``negative" nearest and next to the nearest neighbors 
couplings; that is, starting from a model which is in some sense ferromagnetic, 
we obtain a model with antiferromagnetic couplings.
\par
The phase diagram of the eight vertex model is well known (see, e. g., 
\cite{BAX,NL1,L}): at $J_{<<>>}<0$ and $|J_{<>}|<2\; |J_{<<>>}|$ this phase 
diagram is characterized by a critical surface separating the paramagnetic and 
the super antiferromagnetic (SAF) phases, respectively the high and 
the low temperature phases. We observe that as $\beta$ increases from zero to 
infinity the renormalized eight vertex model (\ref{eq:otto}) undergoes a second 
order phase transition from the paramagnetic to the SAF phase.
\par
Hence, it there exists a value $\beta_c$ such that at $\beta=\beta_c$ our 
original model is ``critical"; the high temperature phase is the usual 
paramagnetic phase: what about the low temperature one? In terms of the new 
variables $s'_{\alpha}\;\alpha\in\Lambda'$ this phase can be characterized by a 
staggered magnetization $m'$ defined as the difference between the 
magnetization on even and odd columns: $m'=0$ in the paramagnetic phase, 
$m'\not= 0$ in the SAF phase. But a column of the new model corresponds to a 
strip of length three lattice spacings in model (\ref{eq:ham}); then, at low 
temperature, model (\ref{eq:ham}) exhibits an ordered phase made of an 
alternate sequence of parallel pluses and minuses polymers, infinitely long and 
large three lattice spacings (see Fig. 2). This suggests that the median roots 
are entropically dominated by polymer--like structures developing parallel
to one of the lattice directions.
\par
We have studied model (\ref{eq:hamti}) following the scheme used for model 
(\ref{eq:ham}); we have partitioned the lattice $\Lambda$ as in Fig. 4. The 
hamiltonian associated to the block variables is
\begin{equation}
-{\tilde H}'(s')= {\rm const}+{\tilde J}_{<>}\sum_{<\alpha\gamma>}s'
 _{\alpha}s'_{\gamma}
 +{\tilde J}_{<<>>}\sum_{<<\alpha\gamma>>}s'_{\alpha}s'_{\gamma}
\label{eq:rinti}
\end{equation}
with ${\tilde J}_{<>}$ and ${\tilde J}_{<<>>}$ depending on $\beta$ as in 
Fig. 5.
\par
Even for model (\ref{eq:hamti}) the renormalized model exhibits a low 
temperature SAF phase. Due to the geometry of the cross blocks, in this case 
the polymers of the low temperature phase are parallel to the dashed lines 
depicted in Fig. 4.

%\newpage
\vskip 2 truecm
\par\noindent
{\bf 4. Conclusions.}
\vskip 0.5 truecm
\par\noindent
We have proposed an Ising--like spin model with a new kind of coupling in order 
to point out that a simple request, such as the tendency of each spin to be 
aligned to the majority of its neighbors, leads to low temperature 
striped--phases.
\par
The expansion of the hamiltonian in terms of products of spin variables pointed 
out the similarities between our model and ANNNI model; this suggests 
the hypothesis that the model 
undergoes a phase transition to a striped--phase. 
\par
We studied the model by Renormalization Group methods and at the level of 
approximation we have considered, first order of cumulant expansion, the phase 
transition is observed.
\par
The  parallel polymerization
appears to be characteristic of a ferromagnetic coupling with threshold.
In the spirit of the ``atomic" interpretation, this implies that the clusters 
of A atoms, as well as B atoms, turn to be one--dimensional at low temperature, 
i. e. they loose one dimension.
\par
Our estimations of critical couplings are 
$\beta_c\sim 25$ for model (\ref{eq:ham}) and $\beta_c\sim 5$ for model 
(\ref{eq:hamti}).
The cumulant expansion is known to be relevant near the critical point. A 
low--temperature series expansion would better describe the model in the limit 
$\beta\to\infty$; this is not a trivial task, since a rigorous analysis of the 
statistical properties of the ground states is needed. This will be the topic 
of further work.

%\newpage
\vskip 2 truecm
\Large
\par\noindent
{\bf Acknowledgements}
\normalsize
\addtolength{\baselineskip}{\baselineskip}
\vskip 1 truecm
\par\noindent
We like to acknowledge L. Angelini, D. Caroppo, G. M. Cicuta, G. Gonnella, 
G. Nardulli, E. Olivieri, I. Sardella and M. Villani for useful discussions.

\newpage
\Large
\par\noindent
{\bf Figure Captions}
\normalsize
\addtolength{\baselineskip}{\baselineskip}
\vskip 1 truecm
\par\noindent
Figure 1: In Fig. 1a the smallest cluster invariant to the transformation 
(\ref{eq:maj}) is shown: black squares represent plus spins on a minuses 
background or, viceversa, minus spins on a pluses background. 
In Fig. 1b a portion of a polymer is depicted. The 
polymer is supposed to be infinitely long; it satisfies the constraint 
(\ref{eq:constr}), as well. 
\vskip 0.5 truecm
\par\noindent
Figure 2: The typical low temperature pattern for our model. The average 
distance between polymers of the same sign is six lattice spacings. The 
constraint (\ref{eq:constr}) is almost satisfied.
\vskip 0.5 truecm
\par\noindent
Figure 3: Renormalized couplings as functions of $\beta$ in the case of model 
(\ref{eq:ham}). Solid, dashed and dotted lines represent respectively $J_{<>}$, 
$J_{<<>>}$ and $J_{
{\vcenter{\vbox{\hrule height.3pt
	    \hbox{\vrule width.3pt height0.15truecm \kern0.15truecm
	    \vrule width.3pt}
	    \hrule height.3pt}}}
}$.
\vskip 0.5 truecm
\par\noindent
Figure 4: The partition of the lattice used to study model (\ref{eq:hamti}).
The renormalized variables are defined on the blank circles, which form the 
new lattice.
\vskip 0.5 truecm
\par\noindent
Figure 5: Renormalized couplings as functions of $\beta$ in the case of model 
(\ref{eq:hamti}). Solid and dashed lines represent respectively 
${\tilde J}_{<>}$ and ${\tilde J}_{<<>>}$.

\newpage
%\vskip 2 truecm
\Large
\par\noindent
{\bf Table Captions}
\normalsize
\addtolength{\baselineskip}{\baselineskip}
\vskip 1 truecm
\par\noindent
{\bf Table 1:} The coefficients $\gamma_i$, which are present in equation 
(\ref{eq:hamff}), are listed. In the third column the families of clusters 
$\Gamma_i$, to which each coefficient $\gamma_i$ is related, are briefly 
described. In the fourth column a typical cluster $\zeta\in\Gamma_i$ is 
depicted: the grid represents the $3\times 3$ block $B$ in which $\zeta$ is 
contained, the sites of $B$ belonging to $\zeta$ are represented by the 
black circles. We remark that just an example of cluster belonging to 
$\Gamma_i$ is depicted in the fourth column; for example, in the case $i=10$ 
one can consider the following clusters, as well:
\begin{picture}(12,12)
\put(0,0){\line(0,12){12}}
\put(6,0){\line(0,12){12}}
\put(12,0){\line(0,12){12}}
\put(0,0){\line(12,0){12}}
\put(0,6){\line(12,0){12}}
\put(0,12){\line(12,0){12}}
\put(0,0){\circle*{3}}
\put(0,6){\circle*{3}}
\put(12,12){\circle*{3}}
\put(12,6){\circle*{3}}
\end{picture}
,
\begin{picture}(12,12)
\put(0,0){\line(0,12){12}}
\put(6,0){\line(0,12){12}}
\put(12,0){\line(0,12){12}}
\put(0,0){\line(12,0){12}}
\put(0,6){\line(12,0){12}}
\put(0,12){\line(12,0){12}}
\put(0,0){\circle*{3}}
\put(0,6){\circle*{3}}
\put(6,12){\circle*{3}}
\put(12,6){\circle*{3}}
\end{picture}
.

\newpage
\thispagestyle{empty}
\Large
\begin{center}
{\bf Table 1}
\end{center}
\normalsize
\vskip 1 truecm
\begin{center}
\begin{tabular}{c|c|p{10truecm}|c} 
\hline\hline
i & $\gamma_i$ & \multicolumn{1}{c|}{description of $\Gamma_i$} & 
\multicolumn{1}{c}{typical cluster in $\Gamma_i$} \\
\hline\hline
1 & +200 & pairs of nearest neighbors sites &
\begin{picture}(12,12)(0,2)
\put(0,0){\line(0,12){12}}
\put(6,0){\line(0,12){12}}
\put(12,0){\line(0,12){12}}
\put(0,0){\line(12,0){12}}
\put(0,6){\line(12,0){12}}
\put(0,12){\line(12,0){12}}
\put(0,6){\circle*{3}}
\put(6,6){\circle*{3}}
\end{picture}
\\
\hline
2 & +240 & pairs of next to the nearest neighbors sites & 
\begin{picture}(12,12)(0,2)
\put(0,0){\line(0,12){12}}
\put(6,0){\line(0,12){12}}
\put(12,0){\line(0,12){12}}
\put(0,0){\line(12,0){12}}
\put(0,6){\line(12,0){12}}
\put(0,12){\line(12,0){12}}
\put(0,6){\circle*{3}}
\put(6,0){\circle*{3}}
\end{picture}
\\
\hline
3 & -60 & pairs of second--neighbor sites along the lattice directions & 
\begin{picture}(12,12)(0,9)
\put(0,0){\line(0,12){12}}
\put(6,0){\line(0,12){12}}
\put(12,0){\line(0,12){12}}
\put(0,0){\line(12,0){12}}
\put(0,6){\line(12,0){12}}
\put(0,12){\line(12,0){12}}
\put(0,6){\circle*{3}}
\put(12,6){\circle*{3}}
\end{picture}
\\
\hline
4 & -40 & 2--sites clusters with sites at distance $\sqrt 5$ lattice spacings & 
\begin{picture}(12,12)(0,2)
\put(0,0){\line(0,12){12}}
\put(6,0){\line(0,12){12}}
\put(12,0){\line(0,12){12}}
\put(0,0){\line(12,0){12}}
\put(0,6){\line(12,0){12}}
\put(0,12){\line(12,0){12}}
\put(0,0){\circle*{3}}
\put(12,6){\circle*{3}}
\end{picture}
\\
\hline
5 & -20 & pairs of second--neighbor sites along the lattice diagonals & 
\begin{picture}(12,12)(0,2)
\put(0,0){\line(0,12){12}}
\put(6,0){\line(0,12){12}}
\put(12,0){\line(0,12){12}}
\put(0,0){\line(12,0){12}}
\put(0,6){\line(12,0){12}}
\put(0,12){\line(12,0){12}}
\put(0,0){\circle*{3}}
\put(12,12){\circle*{3}}
\end{picture}
\\
\hline
6 & -80 & plaquettes & 
\begin{picture}(12,12)(0,2)
\put(0,0){\line(0,12){12}}
\put(6,0){\line(0,12){12}}
\put(12,0){\line(0,12){12}}
\put(0,0){\line(12,0){12}}
\put(0,6){\line(12,0){12}}
\put(0,12){\line(12,0){12}}
\put(0,0){\circle*{3}}
\put(0,6){\circle*{3}}
\put(6,0){\circle*{3}}
\put(6,6){\circle*{3}}
\end{picture}
\\
\hline
7 & -40 & 4--sites clusters containing the center of the block $B$ and 
occupying a $3\times 2$ rectangular block& 
\begin{picture}(12,12)(0,9)
\put(0,0){\line(0,12){12}}
\put(6,0){\line(0,12){12}}
\put(12,0){\line(0,12){12}}
\put(0,0){\line(12,0){12}}
\put(0,6){\line(12,0){12}}
\put(0,12){\line(12,0){12}}
\put(0,0){\circle*{3}}
\put(0,6){\circle*{3}}
\put(0,12){\circle*{3}}
\put(6,6){\circle*{3}}
\end{picture}
\\
\hline
8 & -20 & 4--sites clusters containing the center of the block $B$ and 
occupying the whole $B$ & 
\begin{picture}(12,12)(0,9)
\put(0,0){\line(0,12){12}}
\put(6,0){\line(0,12){12}}
\put(12,0){\line(0,12){12}}
\put(0,0){\line(12,0){12}}
\put(0,6){\line(12,0){12}}
\put(0,12){\line(12,0){12}}
\put(0,0){\circle*{3}}
\put(0,6){\circle*{3}}
\put(12,12){\circle*{3}}
\put(6,6){\circle*{3}}
\end{picture}
\\
\hline
9 & +24 & 4--sites clusters not containing the center of the block $B$ and 
occupying a $3\times 2$ rectangular block& 
\begin{picture}(12,12)(0,9)
\put(0,0){\line(0,12){12}}
\put(6,0){\line(0,12){12}}
\put(12,0){\line(0,12){12}}
\put(0,0){\line(12,0){12}}
\put(0,6){\line(12,0){12}}
\put(0,12){\line(12,0){12}}
\put(0,0){\circle*{3}}
\put(0,6){\circle*{3}}
\put(12,6){\circle*{3}}
\put(6,0){\circle*{3}}
\end{picture}
\\
\hline
10 & +12 & 4--sites clusters not containing the center of the block $B$ and 
occupying the whole $B$ & 
\begin{picture}(12,12)(0,9)
\put(0,0){\line(0,12){12}}
\put(6,0){\line(0,12){12}}
\put(12,0){\line(0,12){12}}
\put(0,0){\line(12,0){12}}
\put(0,6){\line(12,0){12}}
\put(0,12){\line(12,0){12}}
\put(0,0){\circle*{3}}
\put(0,6){\circle*{3}}
\put(12,12){\circle*{3}}
\put(12,0){\circle*{3}}
\end{picture}
\\
\hline
11 & +12 & 6--sites clusters containing the center of the block $B$ and 
occupying the whole $B$ & 
\begin{picture}(12,12)(0,9)
\put(0,0){\line(0,12){12}}
\put(6,0){\line(0,12){12}}
\put(12,0){\line(0,12){12}}
\put(0,0){\line(12,0){12}}
\put(0,6){\line(12,0){12}}
\put(0,12){\line(12,0){12}}
\put(0,0){\circle*{3}}
\put(0,6){\circle*{3}}
\put(12,12){\circle*{3}}
\put(6,6){\circle*{3}}
\put(6,0){\circle*{3}}
\put(6,12){\circle*{3}}
\end{picture}
\\
\hline
12 & +24 & 6--sites clusters containing the center of the block $B$ and 
occupying a $3\times 2$ rectangular block & 
\begin{picture}(12,12)(0,9)
\put(0,0){\line(0,12){12}}
\put(6,0){\line(0,12){12}}
\put(12,0){\line(0,12){12}}
\put(0,0){\line(12,0){12}}
\put(0,6){\line(12,0){12}}
\put(0,12){\line(12,0){12}}
\put(0,0){\circle*{3}}
\put(0,6){\circle*{3}}
\put(0,12){\circle*{3}}
\put(6,6){\circle*{3}}
\put(6,0){\circle*{3}}
\put(6,12){\circle*{3}}
\end{picture}
\\
\hline
13 & -20 & 6--sites clusters not containing the center of the block $B$ and, 
necessarily, occupying the whole block $B$& 
\begin{picture}(12,12)(0,9)
\put(0,0){\line(0,12){12}}
\put(6,0){\line(0,12){12}}
\put(12,0){\line(0,12){12}}
\put(0,0){\line(12,0){12}}
\put(0,6){\line(12,0){12}}
\put(0,12){\line(12,0){12}}
\put(0,0){\circle*{3}}
\put(0,6){\circle*{3}}
\put(0,12){\circle*{3}}
\put(12,0){\circle*{3}}
\put(6,0){\circle*{3}}
\put(6,12){\circle*{3}}
\end{picture}
\\
\hline
14 & -20 & 8--sites clusters containing the center of the block $B$ &
\begin{picture}(12,12)(0,2)
\put(0,0){\line(0,12){12}}
\put(6,0){\line(0,12){12}}
\put(12,0){\line(0,12){12}}
\put(0,0){\line(12,0){12}}
\put(0,6){\line(12,0){12}}
\put(0,12){\line(12,0){12}}
\put(0,0){\circle*{3}}
\put(0,6){\circle*{3}}
\put(0,12){\circle*{3}}
\put(6,0){\circle*{3}}
\put(6,6){\circle*{3}}
\put(6,12){\circle*{3}}
\put(12,0){\circle*{3}}
\put(12,6){\circle*{3}}
\end{picture}
\\
\hline
15 & +140 & 8--sites clusters not containing the center of the block $B$ &
\begin{picture}(12,12)(0,2)
\put(0,0){\line(0,12){12}}
\put(6,0){\line(0,12){12}}
\put(12,0){\line(0,12){12}}
\put(0,0){\line(12,0){12}}
\put(0,6){\line(12,0){12}}
\put(0,12){\line(12,0){12}}
\put(0,0){\circle*{3}}
\put(0,6){\circle*{3}}
\put(0,12){\circle*{3}}
\put(6,12){\circle*{3}}
\put(12,12){\circle*{3}}
\put(12,6){\circle*{3}}
\put(12,0){\circle*{3}}
\put(6,0){\circle*{3}}
\end{picture}
\\
\hline\hline
\end{tabular}
\end{center}


\newpage
\begin{thebibliography}{99}
\bibitem{BAX} R. J. Baxter, {\it Exactly Solved Models in Statistical 
Mechanics} (Academic Press, London, 1982). 
\bibitem{YEO} J. Yeomans, {\it Solid State Physics}, Vol. 41 (Academic Press, 
Orlando, 1988).
\bibitem{SEL} W. Selke, in ``Phase Transitions and Critical Phenomena", 
Vol. 15, Eds. C. Domb, J. L. Lebowitz (Academic Press, New York, 1992).
\bibitem{FS} M. E. Fisher and W. Selke, Phys. Rev. Lett. {\bf 44}, 1502 (1980).
\bibitem{GRIF} R. B. Griffiths, Journal of Mathematical Physics {\bf 8}, n. 6, 
478 (1967) and R.B. Griffiths, Journal of Mathematical Physics {\bf 8}, n. 6, 
484 (1967).
\bibitem{ma} Sheng--Keng Ma, {\it Modern Theory of Critical Phenomena} (W. A. 
Benjamin, Inc., 1976).
\bibitem{NL1} Th. Niemeijer, J. M. J. van Leeuwen, in ``Phase Transitions and 
Critical Phenomena", Vol. 6, Eds. C. Domb, M. S. Green (Academic Press, New 
York, 1976).
\bibitem{KH} K. Huang, {\it Statistical Mechanics} (John Wiley \& Sons, 1987).
\bibitem{TSH} T. S. Huang, {\it Two--Dimensional Digital Signal Processing II} 
(Springer--Verlag, 1981).
\bibitem{DER} B. Derrida, in {\it Fundamental Problems in Statical Mechanics 
VII} (Elsevier Science Publishers B.V., 1990).
\bibitem{NL} Th. Niemeijer, J. M. J. van Leeuwen, Physica {\bf 71}, 17 (1974);
Th. Niemeijer, J. M. J. van Leeuwen, Phys. Rev. Lett. {\bf 31}, 1411 (1973).
\bibitem{L} J. M. J. van Leeuwen, Phys. Rev. Lett. {\bf 34}, 1056 (1975).
\end{thebibliography}
\end{document}